# Stratification of carotid atheromatous plaque using interpretable deep learning methods on B-mode ultrasound images

Theofanis Ganitidis, Maria Athanasiou, Kalliopi Dalakleidi, *Member, IEEE*, Nikos Melanitis, Spyretta Golemati, *Member, IEEE*, and Konstantina S. Nikita, *Fellow, IEEE*

*Abstract*— Carotid atherosclerosis is the major cause of ischemic stroke resulting in significant rates of mortality and disability annually. Early diagnosis of such cases is of great importance, since it enables clinicians to apply a more effective treatment strategy. This paper introduces an interpretable classification approach of carotid ultrasound images for the risk assessment and stratification of patients with carotid atheromatous plaque. To address the highly imbalanced distribution of patients between the symptomatic and asymptomatic classes (16 vs 58, respectively), an ensemble learning scheme based on a sub-sampling approach was applied along with a two-phase, cost-sensitive strategy of learning, that uses the original and a resampled data set. Convolutional Neural Networks (CNNs) were utilized for building the primary models of the ensemble. A six-layer deep CNN was used to automatically extract features from the images, followed by a classification stage of two fully connected layers. The obtained results (Area Under the ROC Curve (AUC): 73%, sensitivity: 75%, specificity: 70%) indicate that the proposed approach achieved acceptable discrimination performance. Finally, interpretability methods were applied on the model's predictions in order to reveal insights on the model's decision process as well as to enable the identification of novel image biomarkers for the stratification of patients with carotid atheromatous plaque.

*Clinical Relevance*—The integration of interpretability methods with deep learning strategies can facilitate the identification of novel ultrasound image biomarkers for the stratification of patients with carotid atheromatous plaque.

I. INTRODUCTION

Cardiovascular Diseases (CVDs), a group of diseases involving the heart and the blood vessels, are the leading cause of death and disability worldwide. Over 9.5 million new cases of ischemic stroke occur annually worldwide, with 60% of them occurring in people under 70 years. The cost of such rates includes 2.7 million deaths and 51.9 million years of healthy life lost annually due to ischemic stroke related mortality and disability [1], [2]. An important part of CVDs and stroke is related to atherosclerosis.

Atherosclerosis is a chronic and progressive disease characterized by the accumulation of lipids and fibrous tissue within the arterial wall. In general, carotid atheromatous plaque is related to restriction or disruption of blood flow to the brain and is the main cause of ischemic stroke. It usually remains asymptomatic for several years and is therefore hard to detect. An early and accurate detection of individuals who are vulnerable to cardiovascular events or at high risk of stroke would help physicians plan a therapeutic strategy and apply preventive measures before the aforementioned life-threatening events take place.

Carotid sonography plays a vital role for the early diagnosis of such conditions since it provides the possibility of risk assessment based on both qualitative and quantitative features[2]-[1]. Due to its low cost, wide availability and non-invasive nature, ultrasound has the potential to become the modality of choice for plaque detection in clinical practice.

However, the risk prediction and the identification of patients at high risk is a non-trivial and non-formulated issue. Some risk predictors are proposed for the identification of individuals who are at high risk, by evaluating some major risk factors, such as age, gender, hypercholesterolemia, hypertension, body mass index, family history, smoking and diabetes [5]. Factors like shear stress at the vessel wall or blood flow velocity are also critical factors for the formation and the development of plaque. Although these factors clearly correlate with the extent of plaque formation, it is unclear how they influence the vulnerability of plaques. These predictions, calculated by formulated risk factors, do not provide patient-specific information, thereby resulting in inefficient estimations of atherosclerosis risk. Recent developments in deep learning and image analysis hold promise for providing the necessary and detailed features [6]-[9].

Deep learning is successfully used as a tool for data-driven machine learning, where a multilayered neural network is trained on a dataset to automatically learn distinctive features. Among deep learning techniques, deep convolutional networks are used for the purpose of medical image analysis. The growing availability of digital images related to variable clinical tasks, combined with the improvement of computational systems, has made it possible to develop efficient and accurate diagnostic models [9], [10].

Recently, deep learning medical image analysis has given promising results in patient stratification [9], [10]. However, one implicit limitation of deep neural networks is the black box operation principle. Clinical applications require high

T. G., M. A., K. D., N. M., S. G., K. S. N. are with the Biomedical Simulations and Imaging (BIOSIM) Laboratory, National Technical University of Athens (NTUA), 9, Iroon Polytechniou Str., 15780 Zografos, Athens Greece. (corresponding author e-mail: theogani@biosim.ntua.gr). S.G. is also with the Medical School, National and Kapodistrian University of Athens, Athens, Greece.

level of transparency. This means that we need to be able to explain system decisions and predictions and justify their reliability in order to include them in clinical practice. Except for the transparency of the decision-making process, it is also beneficial to exploit knowledge from black box prediction models and try to investigate possible causal relationships between previously unknown features.

Following the aforementioned demands, a series of methods, known as interpretable artificial intelligence, have been developed [11], [12]. These methods aim either to develop models with implicit interpretability of their operation, or to develop tools which add post-hoc interpretability to a pre-existing black-box model. The former use algorithms like K-Nearest Neighbors, Decision Trees or Rule-based Learners with interpretable predictions. The latter use a series of model-specific or model agnostic methods that aim to indirectly explain the predictions of the underlying black box model. In this work, we focus on leveraging the unique feature extraction capabilities of deep Convolutional Neural Networks (CNNs) along with State-of-the-Art interpretability methods towards the development of an interpretable model for the risk stratification of patients with carotid atheromatous plaque.

## II. METHODS

### A. Dataset

The dataset used includes sequences of B-mode ultrasound images (videos) of patients referred to Attikon General University Hospital of Athens for carotid ultrasonography. The local institutional review board approved the study protocol and all subjects gave their informed consent to the scientific use of the data. A total of 12 symptomatic and 41 asymptomatic patients with carotid atherosclerosis were included in the study. Symptomatic patients had experienced a stroke or a transient ischemic attack within 6 months prior to the ultrasound examination. Asymptomatic patients presented with no such symptoms. For 3 of the 41 asymptomatic patients, there were two videos available, one from the left and one from the right carotid artery. For 4 symptomatic and 14 asymptomatic patients, there were two plaques in the ultrasound videos. Thus, a total of 74 videos were investigated, 58 corresponding to asymptomatic and 16 to symptomatic cases. For each case the atheromatous plaque was cropped according to the radiologist's segmentation.

### B. Convolutional Neural Network Classification

Convolutional neural networks (CNNs) are a kind of neural network architecture that present some valuable features, relevant for the image representation task. Each layer of a CNN uses a group of trainable filters which are convoluted over the input image. This parameter-sharing feature makes it possible to extract high-level, abstract image features by stacking a number of CNN layers. Thus, CNNs have shown their promise by successfully addressing several challenging image recognition tasks [13].

For the classification problem under study, the architecture, illustrated in Fig. 1, was implemented. For the feature extraction, we implemented a design consisting of six convolutional layers with kernel sizes of 5x5, each followed by rectified linear units (ReLU), three average pooling layers with pool size of 3x3. For the classification stage we used a fully connected layer with ReLU activation followed by a softmax fully connected layer. We used the binary cross entropy loss function. As a regularization scheme we used L2 penalty on the classification layers, dropout layers after the convolutional and fully connected layers at training phase, as well as early stopping. We solved the optimization problem using the Adam method [14]. The regularization hyperparameters were validated using a 4-fold cross validation scheme to select the best combination.

### C. Ensemble Approach – Combination Schemes

As stated previously, in our dataset, the symptomatic class is significantly underrepresented. This is a common issue in clinical datasets. To overcome this problem, we applied an ensemble learning method, where three primary models with identical architectures were trained individually on different parts of the training set [15]. We also used two-phase training and cost-sensitive learning [16][17].

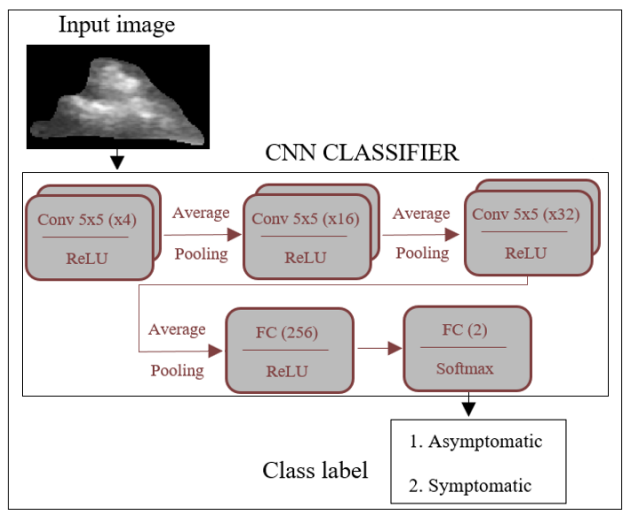

*Figure 1: Schematic diagram of the CNN architecture used for the primary models. It consisted of six convolutional (Conv), three average pooling and two fully connected (FC) layers. Softmax activation function was used for the top FC layer and rectified linear units (ReLU) for the rest of the layers.*

Firstly, we randomly split the dataset into training, validation and testing sets, in proportions of 62%, 13% and 25%, respectively. The validation and testing sets remained unaltered for the entire procedure, so as to retain the original distribution of patients in the two classes. Subsequently, a balanced sub-sampling approach was adopted, where training sub-sets were generated, preserving a 1:1 ratio between the majority (Asymptomatic) and the minority (Symptomatic) class. To this end, the instances of the majority class in the initial training dataset were firstly divided into three parts; and each part was merged with all the instances of the minority class towards the creation of three training sub-sets. Each sub-set was used for training a CNN-based primary model. Fig. 2 illustrates schematically the described process.

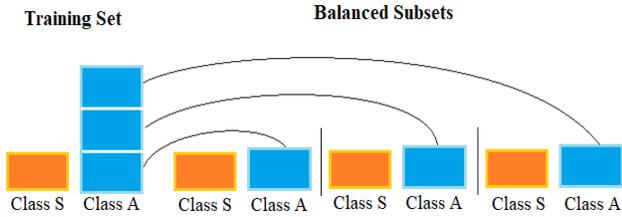

Figure 2: Training set division into three balanced subsets.

For of each of the CNN-based primary models, a two-phase training was applied. In the first phase, each primary model was trained on one of the sub-sets described above. Next, we applied a fine-tuning step, where the fully connected layers were trained on the original, imbalanced training set for several epochs. Cost-sensitive learning was deployed and a weighted version of the loss function was used by introducing a weight factor to enforce the learning of positive (Symptomatic) instances. During this phase, the training hyperparameters of the first phase were applied.

Three schemes were investigated for combining the outputs of the primary models. The first was a voting scheme. According to this, each primary model's output is transformed to a vote by applying a probability threshold, and the final output of the ensemble is the class with the majority of votes. The second scheme was based on the probabilities given by each individual primary model. In this approach, the probability of an image to belong to a certain class was derived as the average of the probabilities of the primary models. Finally, the third combination scheme was a weighted average scheme. In this approach, the probabilities of each individual primary classifier were weighted before being averaged to calculate the final output of the ensemble. The weights were set proportional to the certainties ($c_i$) of the respective primary model output, using the following equations:

$$w_i = \frac{c_i}{\sum_{i=1}^{3} c_i} \quad (1)$$

$$c_i = \begin{cases} y_i \text{ if } y_i \geq 0.5 \\ 1 - y_i \text{ if } y_i < 0.5 \end{cases} \quad (2)$$

where $w_i$ is the weight of the i-th primary model, $c_i$ is the certainty of the i-th primary model and $y_i$ is the probability given by the i-th primary model for the input to belong to the positive class.

TABLE I. AVERAGE EVALUATION METRICS OF THE 4-FOLD CROSS VALIDATION

| Evaluation metrics | Sensitivity | Specificity | Balanced Accuracy | AUC |
|---|---|---|---|---|
| 4-Fold Average | 75±17.6% | 70±10.3% | 72.5±6% | 73±10.7% |

### D. Interpretability methods

In this work we used local surrogate models to explain the individual predictions of the neural network. Local surrogate models are interpretable models which are trained to approximate the predictions of the underlying black box model [18]. The idea is that we have a data instance and we aim to explain the prediction of a black box model on this instance. To do so, we get perturbations of this instance and we train an interpretable model to locally approximate the prediction of the black box model. Image perturbations are defined as follows. Firstly, unsupervised segmentation of the image into super pixels is performed. Then, super pixels are randomly replaced by a specified gray-scale value. The resulting images with randomly hidden super pixels are the perturbations of the original image. In this way, the interpretable local surrogate model is able to quantify the impact of each super pixel on the prediction. Since the interpretable model is trained to approximate our model locally, it can indicate which super pixels impact drastically the prediction and whether they have a positive or negative impact.

### D. Evaluation metrics

To assess the generalization ability of the ensemble models, 4-fold cross-validation was used. Accuracy, sensitivity, specificity, balanced accuracy and area under the ROC curve (AUC) were used in order to evaluate the model's discrimination ability.

## III. EXPERIMENTAL SETUP AND RESULTS

### A. Hyperparameter Validation

The classifier's parameters, including kernel sizes and the number of filters of each CNN layer and the number of neurons in the fully connected layers, were validated using its 4-fold cross-validation score to select the best setting amongst several combinations. Moreover, the regularization parameters, including dropout probabilities and L2 regularization hyperparameter (lambda), were also selected by optimizing the evaluation score in a 4-fold cross-validation manner.

Three combination schemes were also evaluated as mentioned above. As far as average and weighted average schemes are concerned, the probability threshold, above which an instance is considered as positive (Symptomatic), was also examined. The result of this process showed that the weighted averaging scheme had a slightly more consistent performance across different parameters' settings of the cross-validation process. However, the best performance was achieved using the simple averaging scheme with a decision threshold of 0.465. The corresponding average values of the evaluation metrics are summarized in Table I. Fig. 3 illustrates a summary confusion matrix.

Figure 3: Confusion matrix.

### B. Prediction interpretation

We examined some of the predictions on the test set to qualitatively evaluate the model's performance. In Fig. 4 we show examples of interpretation for a true negative (a) and a false negative case (b). In the heatmap visualization, the significance of a region is indicated by its absolute value. A negative (red) value indicates that the corresponding region impacts the prediction towards the asymptomatic class and vice versa. The juxtaluminal hypoechoic area (JBA), that is a well-established criterion for the risk assessment of symptomatic patients [19]0, is also indicated by purple color in the left panels of Fig. 4a and Fig. 4b.

In Fig. 4a an example of an asymptomatic plaque is presented. The areas marked on the ultrasound image are the two most important ones. The red (green) color defines a region which is important for the asymptomatic (symptomatic) class. This means that, with the corresponding area hidden, the final prediction would be affected significantly. Since the image was correctly classified as asymptomatic, we can infer that the green area, despite having a great impact, did not determine the final prediction overall. It should be noted that the green area contains the entire JBA, which is consistent with the literature about the JBA's predictive importance for symptomatic patients.

In Fig. 4b, an example of a symptomatic plaque is presented. Again, it is shown that the green area where the JBA is located plays a key role, driving the model's prediction towards symptomatic class. However, the red region, which leads to the calculation of a lower probability of asymptomaticity, has greater impact as it has greater absolute value. Consequently, this case was misclassified as asymptomatic.

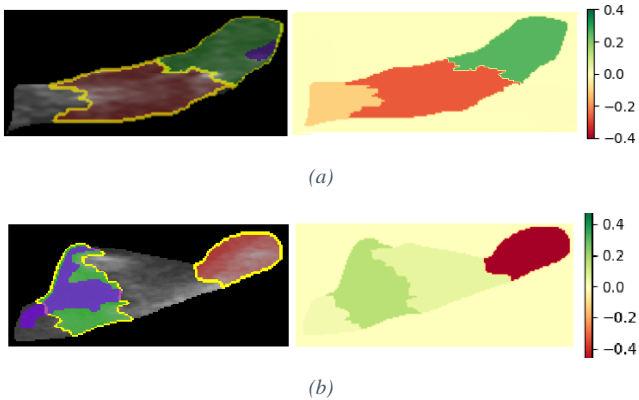

*(a)*

*(b)*

*Figure 4: Prediction interpretation examples. In the heatmap, the absolute value of a region indicates its significance. A positive value indicates that the corresponding region impacts the prediction towards the symptomatic class whereas a negative one impacts towards the asymptomatic. The two most important regions (greatest absolute value) and their borders are also highlighted on the ultrasound as red or green areas according to the sign of their value. The JBA is marked with purple color. (a) Correctly classified asymptomatic case. (b) Missclassified symptomatic case.*

## IV. CONCLUSION

This work investigated a highly imbalanced dataset and introduced an interpretable deep learning approach, based on an ensemble learning method, for the risk stratification of patients with carotid atheromatous plaque. A number of adjustments are expected to improve the overall performance of the proposed model and boost its generalization ability. Such adjustments include the optimization of model architecture, the introduction of more complex resampling strategies, such as targeted resampling or oversampling of the minority class by generating new synthetic data, and the combination with different modalities (multimodal learning), such as biochemical markers related to the risk assessment. Future work includes evaluation on a larger dataset and examination of the special features that affect the discrimination between the two classes (i.e., the features that determine whether a plaque is vulnerable or not). As indicated in the presented examples, the significant areas may highlight an unknown underlying correlation between morphological features of the plaque and the related risk of plaque rupture which should be further examined.